\documentclass[%
aip,
 amsmath,amssymb,
 reprint,%
]{revtex4-1}

\usepackage{graphicx}
\usepackage{dcolumn}
\usepackage{bm}

\usepackage[utf8]{inputenc}
\usepackage[T1]{fontenc}
\usepackage{mathptmx}
\usepackage{etoolbox}
\usepackage[colorlinks,citecolor=blue]{hyperref}
\usepackage{float}
\makeatletter
\def\@email#1#2{%
 \endgroup
 \patchcmd{\titleblock@produce}
  {\frontmatter@RRAPformat}
  {\frontmatter@RRAPformat{\produce@RRAP{*#1\href{mailto:#2}{#2}}}\frontmatter@RRAPformat}
  {}{}
}%
\makeatother
\begin{document}

	\title{Flavor symmetric origin of texture zeros in minimal inverse seesaw and impacts on leptogenesis}
	\author{Nayana Gautam}
	
	\author{Mrinal Kumar Das}
	\affiliation{Department of Physics, Tezpur University, Tezpur - 784028, India}

	
	\begin{abstract}
		We study the prediction of maximal zeros of the Dirac mass matrix on neutrino phenomenology and baryon asymmetry of universe (BAU) within the framework of an inverse seesaw ISS $(2,3)$. We try to find the origin of the allowed two zero textures of the Dirac mass matrix from $S_{4}$ flavor symmetry. ISS $(2,3)$ model contains two pairs of quasi-Dirac particles and one sterile state in the keV scale along with three active neutrinos. The decays of the quasi-Dirac pairs create lepton asymmetry that can be converted to baryon asymmetry of the universe by the sphaleron process. Thus, BAU can be explained in this framework through leptogenesis. We study BAU in all the two zero textures of the Dirac mass matrix. The viabilities of the textures within the framework have been verified with the latest cosmology data on BAU.
		 
	\end{abstract}
	
	\maketitle
	
	\section{\label{sec:1}Introduction:}
	Baryon asymmetry of the
	universe (BAU) has been an appealing mystery in particle physics as well as cosmology. \cite{Weinberg:1979sa,Sakharov:1967dj,kolb2018early}. There are robust observational evidences of BAU that was produced after the Big Bang, but the origin is still an open question to the physics community. The quantitative measurement of the asymmetry can be expressed as the baryons to photon ratio 
	\begin{equation}
	\eta_{B}= \frac{n_{B}- n_{\bar{B}}}{n_{\gamma}}
	\end{equation}
	where $n_{B}$, $n_{\bar{B}}$ and $n_{\gamma}$ are the baryon number density, anti baryon number density and photon density respectively.
	Another expression in terms of entropy can be obtained as,
	\begin{equation}
	Y_{B}= \frac{n_{B}- n_{\bar{B}}}{s}
	\end{equation}
	 Current cosmological bounds for the baryon asymmetry are \cite{Planck:2018vyg}, 
	\begin{equation}
	\eta_{B}=(6.1 \pm 0.18)\times 10^{-10}, \hspace{2mm} Y_{B}=(8.75 \pm 0.23)\times 10^{-11}
	\end{equation} 
	Among several beyond standard model frameworks, the inverse seesaw mechanism can provide a way to explain BAU along with neutrino phenomenology at a very low scale compared to the conventional seesaw \cite{MA1987287,Wyler:1982dd,Mohapatra:1986bd,Deppisch:2004fa,Dev:2009aw}. In inverse seesaw ISS (2,3), lepton asymmetry can be produced through out of equilibrium decay of the quasi-Dirac pairs present in the model \cite{Dev2010,Blanchet_2010,abada2017neutrino,cosmo1}. The Yukawa couplings of the heavy neutrinos are the sole source of CP violation in this mechanism. The leptonic asymmetry gets resonantly enhanced as the RH neutrino mass spectrum has a certain amount of degeneracy comparable to their decay width in this model \cite{Fukugita:1986hr,Pilaftsis:2003gt,PILAFTSIS_1999}. Leptogenesis can provide a highly nontrivial link between baryon asymmetry of the universe and neutrino masses and mixings. Again, the free parameters in the model can be reduced by implementing texture zeros to the mass matrices involved in the model \cite{Ghosal:2015lwa,Kageyama:2002zw,Branco:2007nb,Choubey:2008tb,Adhikary:2013mfa,Sinha:2015ooa,Meloni:2014yea,Alcaide:2018vni,Fritzsch:2011qv,Grimus:2004hf,Samanta:2015oqa}. As mentioned in our earlier work \cite{Gautam:2021vhf}, there are 12 possible two zero textures in ISS(2,3) that can be divided into six classes. In this paper, we try to find the flavor symmetric origin of the mass matrices involved in this model. After constructing the zero textures of $M_D$, we perform a detailed analysis of baryon asymmetry created in all the textures. The viability of the different textures is identified by implementing astrophysical and cosmological bounds on baryon asymmetry of the universe.
	
	The paper is structured as follows. In section \ref{sec:2}, we have shown the $S_4$ flavor symmetric origin of possible texture zeros in the framework of ISS $(2,3)$ along with the description of the framework. In section \ref{sec:level4}, we briefly discuss about leptogenesis in ISS $(2,3)$. Section \ref{sec:level6} contains the results of the numerical analysis and discussion. Finally, in section \ref{sec:level8}, we present our conclusion.
	\section{\label{sec:2}$S_{4}$ flavor symmetric origin of texture zeros in Dirac mass matrices Of ISS(2,3)}
	As already mentioned, we have chosen a minimal inverse seesaw model for our study where two RH neutrinos instead of three are included along with the gauge singlet fermions \cite{Abada:2014vea}. The relevant Lagrangian and the expressions for the light neutrino mass matrix can be found in \cite{Abada:2014vea}.
	In ISS(2,3) model, the light neutrino mass matrix can be obtained from three mass matrices $M_{D}$, $M_{N}$, and $\mu$. Therefore, the allowed number of zeros in the light neutrino mass matrix further constrains the structure of these three matrices \cite{Meloni:2014yea}. In our study, we have considered maximal possible zeros in $M_{N}$ and $\mu$ \cite{Camara:2020efq}. The charge assignments of the particles in the model under the symmetry group $S_{4}$ are shown in table \ref{tab1} \cite{Gautam:2019pce}
	\begin{table}[H]
		\centering
		\begin{tabular}{|c|c|c|c|c|c|c|c|c|c|c|c|c|c|c|c|}
			
			\hline 
			Field	& $L$ & $l_{R}$ & $\bar{N_{R}}$ & H & s & $\phi$ &  $\phi^{\prime}$ &$\phi_{s}$ & $\phi_{l}$  & $\chi$ & $\chi^{\prime}$ \\ 
			\hline 
			$S_{4}$ &$3_{1}$  & $3_{1}$& $2$ & $1_{1}$ & $1_{1}$& $3_{2}$ & $3_{1}$ &$1_{1}$ & $1_{1}$  &$3_{2}$  & $3_{1}$   \\
			\hline 
			$SU(2)_{L}$ & $2$ & $1$ &$1$& $2$ & $1$ & $1$ &  $1$ &$1$&$1$ & $1$ & $1$  \\
			\hline 
			$Z_{4}$& $1$ & $1$ & $i$ &$1$& $+i$ & $-i$ & $-i$ &  $-1$ &$-i$ & $-i $ & $-i$\\
			\hline 
			$Z_{3}$& $\omega^{2}$ & $1$ &$1$& $1$ & $\omega$ & $\omega$ &  $\omega$ &$\omega$ & $\omega$ & $\omega^{2}$ & $\omega^{2}$  \\
			\hline 
		\end{tabular} 
		\caption{Fields and their respective transformations under the symmetry group of the model.} \label{tab1}
	\end{table}
	The Yukawa Lagrangian for the charged leptons and also for the neutrinos can be expressed as:
	\begin{equation}\label{eq:2}
	-\mathcal{L}  = \mathcal{L}_{\mathcal{M_{L}}}+\mathcal{L}_{\mathcal{M_{D}}} + \mathcal{L}_{\mathcal{M_{N}}}+ \mathcal{L}_{\mathcal{M_{S}}}+ h.c
	\end{equation}
	As mentioned in our work \cite{Gautam:2019pce}, we can obtain the matrices $M_{N}$ and $\mu$ and the charged lepton mass matrix $m_{l}$ as follows,
	\begin{equation}
	M_{N} = \frac{1}{\Lambda}\left(\begin{array}{ccc}
	y_{1r} v_{r}v_{r}^{\prime} & 0 & 0 \\
	0 & y_{2r}v_{r}^{\prime}v_{r}^{\prime} & 0\\
	\end{array}\right).
	\end{equation}
	\begin{equation}
	\mu  = y_{s}\left(\begin{array}{ccc}
	1 & 0 & 0 \\
	0 & 1 & 0\\ 
	0 & 0 & 1
	\end{array}\right)v_{s}
	\end{equation}
	\begin{equation}\label{eq:g}
	m_{l}= \frac{v_{h}}{\Lambda}\left(\begin{array}{ccc}
	y_{l}v_{l} & 0& 
	0 \\
	0 & y_{l}v_{l} &
	0\\ 
	0& 0 & y_{l}v_{l}
	\end{array}\right),
	\end{equation}
	Here, we define $y_{s}v_{s}= p$ ,$y_{1r} v_{r}v_{r}^{\prime}= f$ and $y_{2r}v_{r}^{\prime}v_{r}^{\prime}= g$. 
	With these structures of $M_{N}$ and $\mu$, the allowed two zero texture zeros of $M_{D}$ have already been shown in \cite{Gautam:2021vhf}. However, in this work, we find the flavor symmetric origin of these two zero textures of $M_{D}$ \cite{king2013neutrino,altarelli2010discrete,ishimori2010non,flavor}.\\ 
	\textbf{Class A1:}
	The Dirac Lagrangian in this model can be written as,
	\begin{equation}\label{eq:3}
	\mathcal{L}_{\mathcal{M_{D}}} =  \frac{y}{\Lambda}\bar{N_{R}}L H\phi +  \frac{y\prime}{\Lambda}\bar{N_{R}}LH\phi^{\prime},
	\end{equation}
	The cut-off scale $\Lambda$ is needed to lower the mass dimension to $4$. Here, we have also taken extra scalar $\phi$ and $\phi^{\prime}$ with SM Higgs. Now, with $$\langle \Phi \rangle =(v_{h1},v_{h2},0), \; \langle \phi^{\prime} \rangle = (v_{h1^{\prime}},v_{h2^{\prime}},0)$$ we obtain,
	\begin{align}
	\frac{y}{\Lambda} N_{iR}L_{j}H\Phi &=\frac{y}{\Lambda}H [(N_{1R}L_{2}+N_{2R}L_{3}){\Phi_{1}}\nonumber \\& +(N_{1R}L_{3}+N_{2R}L_{1}){\Phi_{2}}+ (N_{1R}L_{1}+N_{2R}L_{2}){\Phi_{3}}]\nonumber \\
	& =\frac{y\prime}{\Lambda}H [(N_{1R}L_{2}+N_{2R}L_{3}){v_{h1}} +(N_{1R}L_{3}+N_{2R}L_{1}){v_{h2}}]
	\end{align}  
	\begin{align}
	\frac{y}{\Lambda} N_{iR}L_{j}H\Phi^{\prime} &=\frac{y}{\Lambda}H [(N_{1R}L_{2}-N_{2R}L_{3}){\Phi_{1}}^{\prime}\nonumber \\& +(N_{1R}L_{3}-N_{2R}L_{1}){\Phi_{2}}^{\prime} + (N_{1R}L_{1}-N_{2R}L_{2}){\Phi_{3}}^{\prime}]\nonumber \\
	& =\frac{y\prime}{\Lambda}H [(N_{1R}L_{2}-N_{2R}L_{3})v_{h1^{\prime}} +(N_{1R}L_{3}-N_{2R}L_{1})v_{h2^{\prime}}]
	\end{align}  
	Thus the structure of $M_{D}$ can be written as,
	\begin{equation}
	M_{D}^{1} = \frac{v_{h}}{\Lambda}\left(\begin{array}{ccc}
	0 &  y v_{h1} + y v_{h1^{\prime}} & y v_{h1} + y v_{h1^{\prime}}  \\
	y v_{h1} - y v_{h1^{\prime}} & 0 & y v_{h1} - y v_{h1^{\prime}}  \\
	\end{array}\right).
	\end{equation}
	We assign, $b = y v_{h1} - y v_{h1^{\prime}}$,  $c = y v_{h1} + y v_{h1^{\prime}}$,  
	$e = y v_{h1} + y v_{h1^{\prime}}$,$h =  y v_{h1} - y v_{h1^{\prime}}$
	\begin{equation} \label{eq:11} 
	M_{D}^{1} = \left[\begin{array}{ccc}
	0 & c& e \\
	b & 0 & h \\ 
	\end{array}\right]
	\end{equation}    
	This structure of $M_{D}^{1}$ along with $M_{D}^{2}$ \cite{Gautam:2021vhf} lead to $1-0$ texture of  $M_{\nu}$ with zero at $(1,2)$ position.
	\begin{equation} \label{eq:12} 
	M_{\nu}= p\left[\begin{array}{ccc}
	k_{1} ^{2} & 0 & k_{1}k_{2} \\
	0 & k_{2} ^{2} & k_{2}k_{3}\\ 
	k_{1}k_{2} & k_{2}k_{3} & k_{1}^{2}+k_{4} ^{2}
	\end{array}\right]
	\end{equation}
	where in $M_{D}^{1}$, $k_{1}= \frac{b}{g}$,$k_{2}= \frac{h}{g}$,$k_{3}= \frac{c}{f}$,$k_{4}= \frac{e}{f}$
	and in $M_{D}^{2}$, $k_{1}= \frac{a}{g}$,$k_{2}= \frac{d}{g}$,$k_{3}= \frac{e}{g}$,$k_{4}= \frac{h}{g}$\\
	\textbf{Class A2:}
	If after SSB, the VEV acquired by the scalars as, $$\langle \Phi \rangle =(0,v_{h2},v_{h3}), \; \langle \phi^{\prime} \rangle = (0,v_{h2^{\prime}},v_{h3^{\prime}})$$ we obtain,
	\begin{align}
	\frac{y}{\Lambda} N_{iR}L_{j}H\Phi &=\frac{y}{\Lambda}H [(N_{1R}L_{2}+N_{2R}L_{3}){\Phi_{1}} \nonumber \\& +(N_{1R}L_{3}+N_{2R}L_{1}){\Phi_{2}} + (N_{1R}L_{1}+N_{2R}L_{2}){\Phi_{3}}]\nonumber \\
	& =\frac{y\prime}{\Lambda}H [(N_{1R}L_{3}+N_{2R}L_{1}){v_{h2}} +(N_{1R}L_{1}+N_{2R}L_{2}){v_{h3}}]
	\end{align}  
	\begin{align}
	\frac{y}{\Lambda} N_{iR}L_{j}H\Phi^{\prime} &=\frac{y}{\Lambda}H [(N_{1R}L_{2}-N_{2R}L_{3}){\Phi_{1}}^{\prime}\nonumber \\& +(N_{1R}L_{3}-N_{2R}L_{1}){\Phi_{2}}^{\prime} + (N_{1R}L_{1}-N_{2R}L_{2}){\Phi_{3}}^{\prime}]\nonumber \\
	& =\frac{y\prime}{\Lambda}H [(N_{1R}L_{3}-N_{2R}L_{1})v_{h2^{\prime}} +(N_{1R}L_{1}-N_{2R}L_{2})v_{h3^{\prime}}]
	\end{align}  
	Thus the structure of $M_{D}$ can be written as,
	\begin{equation}
	M_{D}^{3} = \frac{v_{h}}{\Lambda}\left(\begin{array}{ccc}
	y v_{h3} + y v_{h3^{\prime}} &  0 & y v_{h2} + y v_{h2^{\prime}}  \\
	y v_{h2} - y v_{h2^{\prime}} & y v_{h3} - y v_{h3^{\prime}} & 0  \\
	\end{array}\right).
	\end{equation}
	We assign, $a =\frac{v_{h}}{\Lambda}(y v_{h3} + y v_{h3^{\prime}})$, $b=\frac{v_{h}}{\Lambda}(y v_{h2} - y v_{h2^{\prime}}) $, $d =\frac{v_{h}}{\Lambda}( y v_{h3} - y v_{h3^{\prime}})$,  
	$e = \frac{v_{h}}{\Lambda}(y v_{h2} + y v_{h2^{\prime}})$
	\begin{equation} \label{eq:11c} 
	M_{D}^{3} = \left[\begin{array}{ccc}
	a & 0 & e \\
	b & d & 0 \\ 
	\end{array}\right]
	\end{equation}    
	These two structures of $M_{D}^{3}$ along with $M_{D}^{4}$ \cite{Gautam:2021vhf} lead to $1-0$ texture of  $M_{\nu}$ with zero at $(2,3)$ position.
	\begin{equation} \label{eq:17} 
	M_{\nu}= p\left[\begin{array}{ccc}
	k_{1} ^{2}+ k_{2}^{2} & k_{2}k_{3} & k_{1}k_{4}\\
	k_{2}k_{3} & k_{3}^{2} & 0\\ 
	k_{1}k_{4} & 0 & k_{4}^{2}
	\end{array}\right]
	\end{equation}
	where in $M_{D}^{3}$, $k_{1}= \frac{b}{g}$,$k_{2}= \frac{a}{f}$, $k_{3}= \frac{c}{f}$,$k_{4}= \frac{h}{g}$
	and in $M_{D}^{4}$, $k_{1}= \frac{a}{g}$,$k_{2}= \frac{b}{g}$,$k_{3}= \frac{d}{g}$,$k_{4}= \frac{e}{f}$\\
	
	\textbf{Class A3:}
	If after SSB, the VEV acquired by the scalars as, $$\langle \Phi \rangle =(v_{h1},0,v_{h3}), \; \langle \phi^{\prime} \rangle = (v_{h1^{\prime}},0,v_{h3^{\prime}})$$ we obtain,
	\begin{align}
	\frac{y}{\Lambda} N_{iR}L_{j}H\Phi &=\frac{y}{\Lambda}H [(N_{1R}L_{2}+N_{2R}L_{3}){\Phi_{1}} \nonumber \\&+(N_{1R}L_{3}+N_{2R}L_{1}){\Phi_{2}} + (N_{1R}L_{1}+N_{2R}L_{2}){\Phi_{3}}]\nonumber \\
	& =\frac{y\prime}{\Lambda}H [(N_{1R}L_{2}+N_{2R}L_{3})v_{h1}+ 0 +(N_{1R}L_{1}+N_{2R}L_{2}){v_{h3}}]
	\end{align}  
	\begin{align}
	\frac{y}{\Lambda} N_{iR}L_{j}H\Phi^{\prime} &=\frac{y}{\Lambda}H [(N_{1R}L_{2}-N_{2R}L_{3}){\Phi_{1}}^{\prime}\nonumber \\& +(N_{1R}L_{3}-N_{2R}L_{1}){\Phi_{2}}^{\prime} + (N_{1R}L_{1}-N_{2R}L_{2}){\Phi_{3}}^{\prime}]\nonumber \\
	& =\frac{y\prime}{\Lambda}H [(N_{1R}L_{2}-N_{2R}L_{3})v_{h1^{\prime}} +0 +(N_{1R}L_{1}-N_{2R}L_{2})v_{h3^{\prime}}]
	\end{align}  
	Thus the structure of $M_{D}$ can be written as,
	\begin{equation}
	M_{D}^{5} = \frac{v_{h}}{\Lambda}\left(\begin{array}{ccc}
	y v_{h3} + y v_{h3^{\prime}} &   y v_{h1} + y v_{h1^{\prime}} &  0\\
	0 & y v_{h3} - y v_{h3^{\prime}} & y v_{h1} - y v_{h1^{\prime}}  \\
	\end{array}\right).
	\end{equation}
	We assign, $a =\frac{v_{h}}{\Lambda}(y v_{h3} + y v_{h3^{\prime}} )$, $d=\frac{v_{h}}{\Lambda}(y v_{h3} - y v_{h3^{\prime}}) $, $c =\frac{v_{h}}{\Lambda}(y v_{h1} + y v_{h1^{\prime}})$,  
	$h = \frac{v_{h}}{\Lambda}(y v_{h1} - y v_{h1^{\prime}})$
	
	\begin{equation} \label{eq:13} 
	M_{D}^{5} = \left[\begin{array}{ccc}
	a & c & 0\\
	0 & d & h \\ 
	\end{array}\right]
	\end{equation}    
	These two structures of $M_{D}^{5}$ along with $M_{D}^{6}$\cite{Gautam:2021vhf} lead to $1-0$ texture of  $M_{\nu}$ with zero at $(1,3)$ position.
	\begin{equation} \label{eq:14} 
	M_{\nu}= p\left[\begin{array}{ccc}
	k_{1}^{2} &  k_{1} k_{3} & 0\\
	k_{1} k_{3} & k_{1}^{2}+k_{3} ^{2} & k_{2} k_{4}\\ 
	0 & k_{2} k_{4} & k_{4} ^{2} 
	\end{array}\right]
	\end{equation}
	where in $M_{D}^{5}$, $k_{1}= \frac{b}{g}$,$k_{2}= \frac{d}{g}$, $k_{3}= \frac{c}{f}$,$k_{4}= \frac{e}{f}$
	and in $M_{D}^{6}$, $k_{1}= \frac{a}{g}$,$k_{2}= \frac{d}{g}$,$k_{3}= \frac{c}{f}$,$k_{4}= \frac{h}{f}$\\
	
	\textbf{Class B1:}
	If after SSB, the VEV acquired by the scalars as, $$\langle \Phi \rangle =(v_{h1},v_{h2},v_{h3}), \; \langle \phi^{\prime} \rangle = (v_{h1^{\prime}},v_{h2},v_{h3})$$ we obtain,
	\begin{align}
	\frac{y}{\Lambda} N_{iR}L_{j}H\Phi &=\frac{y}{\Lambda}H [(N_{1R}L_{2}+N_{2R}L_{3}){\Phi_{1}} \nonumber \\&+(N_{1R}L_{3}+N_{2R}L_{1}){\Phi_{2}} + (N_{1R}L_{1}+N_{2R}L_{2}){\Phi_{3}}]\nonumber \\
	& =\frac{y\prime}{\Lambda}H [(N_{1R}L_{2}+N_{2R}L_{3}){v_{h1}} \nonumber \\&+(N_{1R}L_{3}+N_{2R}L_{1}){v_{h2}} +(N_{1R}L_{1}+N_{2R}L_{2}){v_{h3}}]
	\end{align}  
	\begin{align}
	\frac{y}{\Lambda} N_{iR}L_{j}H\Phi^{\prime} &=\frac{y}{\Lambda}H [(N_{1R}L_{2}-N_{2R}L_{3}){\Phi_{1}}^{\prime}\nonumber \\& +(N_{1R}L_{3}-N_{2R}L_{1}){\Phi_{2}}^{\prime} + (N_{1R}L_{1}-N_{2R}L_{2}){\Phi_{3}}^{\prime}]\nonumber \\
	& =\frac{y\prime}{\Lambda}H [(N_{1R}L_{2}+N_{2R}L_{3})v_{h1^{\prime}}\nonumber \\& +(N_{1R}L_{3}-N_{2R}L_{1})v_{h2} +(N_{1R}L_{1}-N_{2R}L_{2})v_{h3}]
	\end{align}  
	Thus the structure of $M_{D}$ can be written as,
	\begin{equation}
	M_{D}^{7} = \frac{v_{h}}{\Lambda}\left(\begin{array}{ccc}
	y v_{h3} + y v_{h3} & y v_{h1} + y v_{h1^{\prime}} & y v_{h2} + y v_{h2}  \\
	y v_{h2} - y v_{h2} & y v_{h3} - y v_{h3} & y v_{h1} - y v_{h1^{\prime}}  \\
	\end{array}\right).
	\end{equation}
	\begin{equation}
	M_{D}^{7} = \frac{v_{h}}{\Lambda}\left(\begin{array}{ccc}
	y v_{h3} + y v_{h3} & y v_{h1} + y v_{h1^{\prime}} & y v_{h2} + y v_{h2}  \\
	0 & 0 & y v_{h1} - y v_{h1^{\prime}}  \\
	\end{array}\right).
	\end{equation}
	We assign, $a =\frac{v_{h}}{\Lambda}(2y v_{h3})$, $c=\frac{v_{h}}{\Lambda}(y v_{h1} + y v_{h1^{\prime}}) $, $e =\frac{v_{h}}{\Lambda}(2 y v_{h2})$,  
	$h = \frac{v_{h}}{\Lambda}(y v_{h1} - y v_{h1^{\prime}})$
	\begin{equation} \label{eq:11d} 
	M_{D}^{7}= \left[\begin{array}{ccc}
	a & c & e \\
	0 & 0 & h \\ 
	\end{array}\right]
	\end{equation}    
	These two structures of $M_{D}6{7}$ and $M_{D}^{8}$\cite{Gautam:2021vhf} lead to texture of  $M_{\nu}$ of the form:
	\begin{equation} \label{eq:22} 
	M_{\nu}= -p \left[\begin{array}{ccc}
	k_{1} ^{2} & k_{1} k_{2} & k_{1} k_{3}\\
	k_{1} k_{2} & k_{2} ^{2} & k_{2} k_{3}\\ 
	k_{1} k_{3} &k_{2} k_{3} & k_{3} ^{2}+k_{4} ^{2}
	\end{array}\right]
	\end{equation}
	where in $M_{D}^{7}$, $k_{1}= \frac{b}{g}$, $k_{2}= \frac{d}{g}$, $k_{3}= \frac{h}{g}$,$k_{4}= \frac{e}{f}$
	and in $M_{D}^{8}$, $k_{1}= \frac{a}{f}$,$k_{2}= \frac{c}{f}$,$k_{3}= \frac{e}{f}$,$k_{4}= \frac{h}{g}$\\
	
	\textbf{Class B2:}
	If after SSB, the VEV acquired by the scalars as, $$\langle \Phi \rangle =(v_{h1},v_{h2},v_{h3}), \; \langle \phi^{\prime} \rangle = (v_{h1},v_{h2},v_{h3^{\prime}})$$ we obtain,
	\begin{align}
	\frac{y}{\Lambda} N_{iR}L_{j}H\Phi &=\frac{y}{\Lambda}H [(N_{1R}L_{2}+N_{2R}L_{3}){\Phi_{1}} \nonumber \\&+(N_{1R}L_{3}+N_{2R}L_{1}){\Phi_{2}} + (N_{1R}L_{1}+N_{2R}L_{2}){\Phi_{3}}]\nonumber \\
	& =\frac{y\prime}{\Lambda}H [(N_{1R}L_{2}+N_{2R}L_{3}){v_{h1}}\nonumber \\& +(N_{1R}L_{3}+N_{2R}L_{1}){v_{h2}} +(N_{1R}L_{1}+N_{2R}L_{2}){v_{h3}}]
	\end{align}  
	\begin{align}
	\frac{y}{\Lambda} N_{iR}L_{j}H\Phi^{\prime} &=\frac{y}{\Lambda}H [(N_{1R}L_{2}-N_{2R}L_{3}){\Phi_{1}}^{\prime}\nonumber \\& +(N_{1R}L_{3}-N_{2R}L_{1}){\Phi_{2}}^{\prime} + (N_{1R}L_{1}-N_{2R}L_{2}){\Phi_{3}}^{\prime}]\nonumber \\
	& =\frac{y\prime}{\Lambda}H [(N_{1R}L_{2}+N_{2R}L_{3})v_{h1}\nonumber \\& +(N_{1R}L_{3}-N_{2R}L_{1})v_{h2} +(N_{1R}L_{1}-N_{2R}L_{2})v_{h3^{\prime}}]
	\end{align}  
	Thus the structure of $M_{D}$ can be written as,
	\begin{equation}
	M_{D}^{9} = \frac{v_{h}}{\Lambda}\left(\begin{array}{ccc}
	y v_{h3} + y v_{h3} & y v_{h1} + y v_{h1} & y v_{h2} + y v_{h2}  \\
	y v_{h2} - y v_{h2} & y v_{h3} - y v_{h3^{\prime}} & y v_{h1} - y v_{h1}  \\
	\end{array}\right).
	\end{equation}
	\begin{equation}
	M_{D}^{9} = \frac{v_{h}}{\Lambda}\left(\begin{array}{ccc}
	y v_{h3} + y v_{h3} & y v_{h1} + y v_{h1} & y v_{h2} + y v_{h2}  \\
	0 & y v_{h3} - y v_{h3^{\prime}} & 0  \\
	\end{array}\right).
	\end{equation}
	We assign, $a =\frac{v_{h}}{\Lambda}(2y v_{h3})$, $c=\frac{v_{h}}{\Lambda}(2 y v_{h1}) $, $e =\frac{v_{h}}{\Lambda}(2 y v_{h2})$,  
	$d =\frac{v_{h}}{\Lambda}(y v_{h3} - y v_{h3^{\prime}})$
	\begin{equation} \label{eq:11e} 
	M_{D}^{9}= \left[\begin{array}{ccc}
	a & c & e \\
	0 & d & 0 \\ 
	\end{array}\right]
	\end{equation}    
	The structure of $M_{\nu}$ arising from $M_{D}^{9}$ and $M_{D}^{10}$\cite{Gautam:2021vhf} as follows:
	\begin{equation} \label{eq:24} 
	M_{\nu}= -p \left[\begin{array}{ccc}
	k_{1} ^{2} & k_{1} k_{2} & k_{1} k_{3}\\
	k_{1} k_{2} & k_{4} ^{2}+k_{2} ^{2} & k_{2} k_{3}\\ 
	k_{1} k_{3} &k_{2} k_{3} & k_{3} ^{2}
	\end{array}\right]
	\end{equation}
	where in $M_{D1}$, $k_{1}= \frac{b}{g}$, $k_{2}= \frac{d}{g}$, $k_{3}= \frac{h}{g}$,$k_{4}= \frac{c}{f}$
	and in $M_{D2}$, $k_{1}= \frac{a}{f}$,$k_{2}= \frac{c}{f}$,$k_{3}= \frac{e}{f}$,$k_{4}= \frac{d}{g}$\\
	
	\textbf{Class B3:}
	If after SSB, the VEV acquired by the scalars as, $$\langle \Phi \rangle =(v_{h1},v_{h2},v_{h3}), \; \langle \phi^{\prime} \rangle = (v_{h1},v_{h2^{\prime}},v_{h3})$$ we obtain,
	\begin{align}
	\frac{y}{\Lambda} N_{iR}L_{j}H\Phi &=\frac{y}{\Lambda}H [(N_{1R}L_{2}+N_{2R}L_{3}){\Phi_{1}}\nonumber \\& +(N_{1R}L_{3}+N_{2R}L_{1}){\Phi_{2}} + (N_{1R}L_{1}+N_{2R}L_{2}){\Phi_{3}}]\nonumber \\
	& =\frac{y\prime}{\Lambda}H [(N_{1R}L_{2}+N_{2R}L_{3}){v_{h1}}\nonumber \\& +(N_{1R}L_{3}+N_{2R}L_{1}){v_{h2}} +(N_{1R}L_{1}+N_{2R}L_{2}){v_{h3}}]
	\end{align}  
	\begin{align}
	\frac{y}{\Lambda} N_{iR}L_{j}H\Phi^{\prime} &=\frac{y}{\Lambda}H [(N_{1R}L_{2}-N_{2R}L_{3}){\Phi_{1}}^{\prime}\nonumber \\& +(N_{1R}L_{3}-N_{2R}L_{1}){\Phi_{2}}^{\prime} + (N_{1R}L_{1}-N_{2R}L_{2}){\Phi_{3}}^{\prime}]\nonumber \\
	& =\frac{y\prime}{\Lambda}H [(N_{1R}L_{2}+N_{2R}L_{3})v_{h1}\nonumber \\& +(N_{1R}L_{3}-N_{2R}L_{1})v_{h2^{\prime}} +(N_{1R}L_{1}-N_{2R}L_{2})v_{h3}]
	\end{align}  
	Thus the structure of $M_{D}$ can be written as,
	\begin{equation}
	M_{D}^{11} = \frac{v_{h}}{\Lambda}\left(\begin{array}{ccc}
	y v_{h3} + y v_{h3} & y v_{h1} + y v_{h1} & y v_{h2} + y v_{h2^{\prime}}  \\
	y v_{h2} - y v_{h2^{\prime}} & y v_{h3} - y v_{h3} & y v_{h1} - y v_{h1}  \\
	\end{array}\right).
	\end{equation}
	\begin{equation}
	M_{D}^{11} = \frac{v_{h}}{\Lambda}\left(\begin{array}{ccc}
	y v_{h3} + y v_{h3} & y v_{h1} + y v_{h1} & y v_{h2} + y v_{h2^{\prime}}  \\
	y v_{h2} - y v_{h2^{\prime}} & 0 & 0  \\
	\end{array}\right).
	\end{equation}
	We assign, $a =\frac{v_{h}}{\Lambda}(2y v_{h3})$, $b =\frac{v_{h}}{\Lambda}(y v_{h2} - y v_{h2^{\prime}})$ ,$c=\frac{v_{h}}{\Lambda}(2cy v_{h1}) $, $e =\frac{v_{h}}{\Lambda}(y v_{h2} + y v_{h2^{\prime}})$  
	\begin{equation} \label{eq:11f} 
	M_{D}^{11}= \left[\begin{array}{ccc}
	a & c & e \\
	b & 0 & 0 \\ 
	\end{array}\right]
	\end{equation}    
	$M_{D}^{11}$ and $M_{D}^{12}$\cite{Gautam:2021vhf} give rise to the following texture of $M_{\nu}$:
	\begin{equation} \label{eq:26} 
	M_{\nu}= -p \left[\begin{array}{ccc}
	k_{1} ^{2}+k_{2} ^{2} & k_{2} k_{3} & k_{2} k_{4}\\
	k_{2} k_{3} & k_{3}^{2} & k_{3} k_{4}\\ 
	k_{2} k_{4} &k_{3} k_{4} & k_{4} ^{2}
	\end{array}\right]
	\end{equation}
	where in $M_{D}^{11}$, $k_{1}= \frac{b}{g}$, $k_{2}= \frac{a}{f}$, $k_{3}= \frac{c}{f}$,$k_{4}= \frac{e}{f}$
	and in $M_{D}^{12}$, $k_{1}= \frac{a}{f}$,$k_{2}= \frac{b}{g}$,$k_{3}= \frac{d}{g}$,$k_{4}= \frac{h}{g}$\\
	
	After constructing the texture zero structures, we have studied neutrino phenomenology and BAU in all the categories.
	\section{\label{sec:level4} Leptogenesis in ISS(2,3) framework}
		
    As mentioned above, the model leads to two pairs of heavy quasi-Dirac particles and a sterile neutrino in keV scale \cite{abada2014dark,awasthi2013neutrinoless,Ankush:2021opd}. Though there are no direct Majorana mass terms for the RH neutrinos, there is possibility of producing lepton asymmetry in this model \cite{Blanchet_2010}. The decays of the heavy quasi-Dirac particles create the lepton asymmetry which in turn converts into baryon asymmetry by a process known as sphaleron \cite{Lucente:2018uaj,Hambye_2012,Choubey_2010,Agashe:2018cuf}. The decays of the two pairs in the model can produce lepton asymmetries. However, the lepton number violating scatterings of the lightest pair will wash out the lepton asymmetries produced by the heavier pair \cite{dev2010tev}. Therefore, in this work, we calculate the lepton asymmetry produced by the lightest quasi-Dirac pair only. The complex Dirac Yukawa couplings are the sole source of CP violation in producing lepton asymmetry \cite{borah2018common}. 
	The two quasi-Dirac RH neutrino pairs are $N_{i,j}$ with masses $M_{i,j}$ where $M_{i}(i=1,2...4)$ denotes the four heavy neutrino mass eigenvalues. The CP asymmetry produced by the decay of $N_{i}$ into any lepton flavor is given by \cite{Covi_1996,Davidson:2008bu},
	\begin{equation}\label{50}
	\epsilon_{i} = \frac{1}{8\pi}\sum_{j\neq i}\frac{Im[(hh^{\dagger})_{ij}^{2}]}{\sum_{\beta}|h_{i\beta}|^{2}}f_{ij}^{\nu}
	\end{equation}
	where, $h_{i\alpha}$ is the effective Yukawa coupling in the diagonal mass basis. $f_{ij}^{\nu}$ can be written as \cite{Dev_2015,Blanchet:2009kk},
	\begin{equation}\label{51}
	f_{ij}^{\nu} \simeq \frac{(M_{j}^{2}-M_{i}^{2})}{(M_{j}^{2}-M_{i}^{2})^{2}+(M_{j}\Gamma_{j}-M_{i}\Gamma_{i})^{2}}
	\end{equation}
	Here, the decay width of the heavy-neutrino $N_{i}$ is represented by  $\Gamma_{i}$.
	$M_{i}$ are the real and positive eigenvalues of the heavy neutrino mass matrix which are grouped into two quasi-Dirac pairs with the mass splitting of order $\mu_{kk}$ $(k=1,2)$.
	The Yukawa couplings in the flavor basis $(y_{i\alpha})$ can be expressed in terms of the Yukawa couplings in the mass basis $(h_{i\alpha})$  as shown in \cite{Blanchet_2010,Gautam:2020wsd}. The calculations of BAU in this model can be found in \cite{Gautam:2020wsd}. Previously, we have studied leptogenesis in a particular $S_{4}$ flavor symmetric model. In this paper, we study the effect of different two zero textures of the Dirac mass matrix on leptogenesis.
	
	\section{\label{sec:level6} Results of Numerical Analaysis and Discussions}
\begin{figure}[H]
		\begin{center}
			\includegraphics[width=0.30\textwidth]{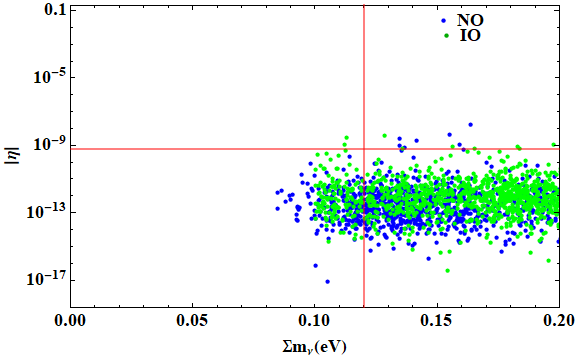}
			
		\end{center}
		\begin{center}
			\caption{BAU as a function of sum of neutrino masses in class A1 for NO and IO.}
			
			\label{fig1}
		\end{center}
	\end{figure}

	The light neutrino mass matrix obtained in all the categories is diagonalised by a unitary PMNS matrix as \cite{giganti2017neutrino},
	\begin{equation}\label{eq:16}
	m_{\nu} = U_{\text{PMNS}}M^{\text{diag}}_{\nu} U^T_{\text{PMNS}}
	\end{equation}
	The diagonal mass matrix of the light neutrinos can be written  as, $M^{\text{diag}}_{\nu} 
	= \text{diag}(0, \sqrt{\Delta m_{solar}^2}, \sqrt{\Delta m_{sol}^2+\Delta m_{atm}^2})$ for normal ordering (NO) and  $M^{\text{diag}}_{\nu} = \text{diag}(\sqrt{\Delta m_{atm}^2}, 
	\sqrt{\Delta m_{solar}^2+ \Delta m_{atm}^2}, 0)$ for inverted ordering (IO) \cite{Nath:2016mts}.
	For the numerical analysis, we have fixed f and g at $9\times10^4$ GeV and $13.5\times10^4$ GeV respectively. Then we have numerically evaluated the other model parameters using \ref{eq:16}. Subsequently, we have calculated the parameters in the light neutrino mass matrix \emph{viz} $k_{1}$, $k_{2}$, $k_{3}$ and $k_{4}$. With the same set of parameters, numerically evaluated for the model, we have calculated the baryon asymmetry in all the six textures. In the calculations, only the decay of the lightest quasi-Dirac pair $(N_{1},s_{1})$ is considered as the asymmetry generated by the heavy pair $(N_{2},s_{2})$ is washed out very rapidly. Thus the same set of model parameters that give correct neutrino phenomenology can also be used for the calculation of the baryon asymmetry of the universe. We also have calculated the sum of the light neutrino masses and have implemented the current cosmological bounds \cite{zyla}.
	Fig \ref{fig1} to fig \ref{fig9} demonstrate the results of our numerical analysis for both normal and inverted ordering.
	Fig \ref{fig1} correlates the baryon asymmetry and the sum of the neutrino masses obtained from the model respectively for NO and IO for the texture A1. It has been observed that A1 leads to the BAU that is consistent with the latest cosmology data. In this texture, the parameter space exceeds the current bound on the sum of the light neutrino masses. However, a small space is available that agrees with both the experimental data on mass squared differences and the current cosmological bound.
	\begin{figure}[H]
		\begin{center}
			\includegraphics[width=0.30\textwidth]{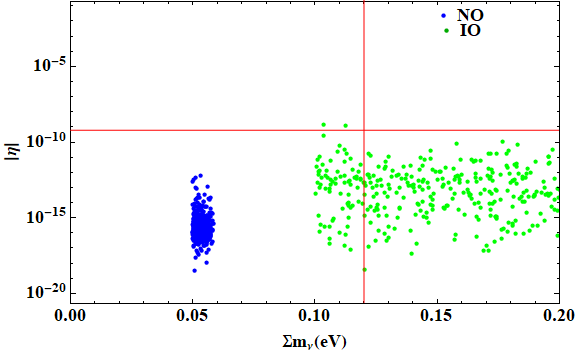}
			
		\end{center}
		\begin{center}
			\caption{BAU as a function of sum of neutrino masses in class A2 for NO and IO.}
			\label{fig2}
		\end{center}
	\end{figure}
	From fig \ref{fig2} , it is clear that the textures A2 do not yield the observed BAU in case of NO. However, the texture A2 in ISS(2,3) can lead to the observed BAU in IO. Thus, the A2 in NO can be discarded from the current cosmological limit on BAU though it gives correct neutrino phenomenology. But a wide range of parameter space of the texture is consistent with the experimental limits on mass squared differences and the cosmological upper bounds in case of IO.
	\begin{figure}[H]
		\begin{center}
			\includegraphics[width=0.30\textwidth]{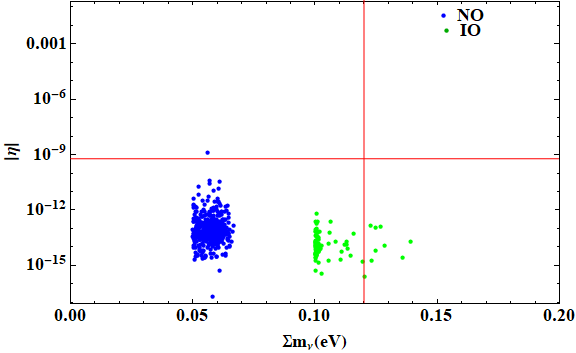}
			
		\end{center}
		\begin{center}
			\caption{BAU as a function of sum of neutrino masses in class A3 for NO and IO.}
			\label{fig3}
		\end{center}
	\end{figure}
	Fig \ref{fig3} shows the baryon asymmetry as a function of the sum of the neutrino masses obtained from the model for NO and IO respectively. We observe that a small parameter space is available that agrees with both the limits in NO. This texture can yield correct amount of BAU in IO as well. 
	\begin{figure}[H]
		\begin{center}
			\includegraphics[width=0.30\textwidth]{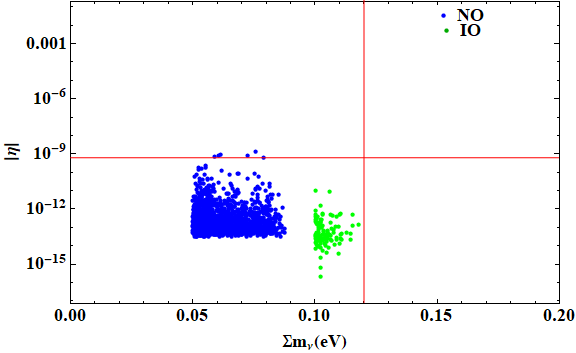}
			
		\end{center}
		\begin{center}
			\caption{BAU as a function of sum of neutrino masses in class B1 for NO and IO.}
			\label{fig4}
		\end{center}
	\end{figure}
	\begin{figure}[H]
		\begin{center}
			\includegraphics[width=0.30\textwidth]{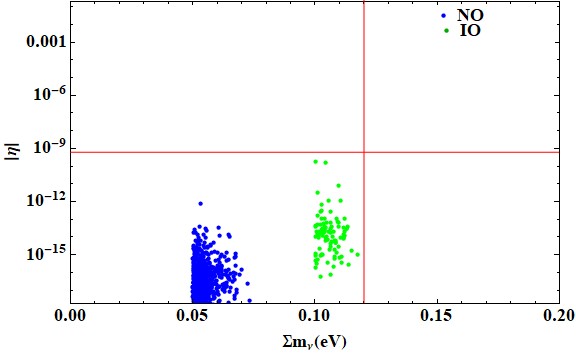}
			
		\end{center}
		\begin{center}
			\caption{BAU as a function of sum of neutrino masses in class B2 for NO and IO.}
			\label{fig5}
		\end{center}
	\end{figure}
Fig \ref{fig4} and fig \ref{fig5} correlate the baryon asymmetry and the sum of the neutrino masses obtained from the model respectively for the textures B1 and B2. It has been observed that for IO, the textures B1 and B2 do not yield the observed BAU. In the case of NO, texture B1 can lead to the correct BAU, But the amount of BAU obtained in B2 is quite less than the observed values. It is seen that the two textures are consistent with the latest cosmology data on the sum of the light neutrino masses and mass squared differences. But the texture B2 is astonishingly weak in their predictions of BAU.
	\begin{figure}[H]
		\begin{center}
			\includegraphics[width=0.30\textwidth]{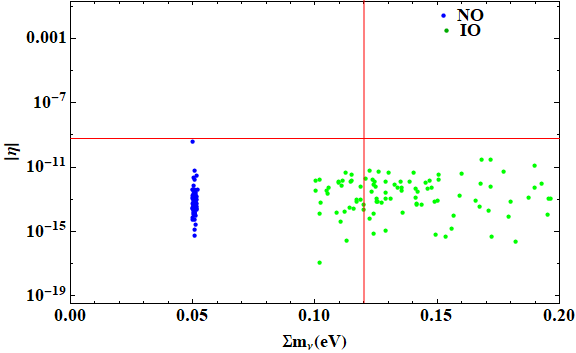}
			
		\end{center}
		\begin{center}
			\caption{BAU as a function of sum of neutrino masses in class B3 for NO and IO.}
			\label{fig6}
		\end{center}
	\end{figure}
Fig \ref{fig6} shows the co relation between baryon asymmetry and the sum of the neutrino masses obtained from the model respectively for NO and IO in texture B3. It has been observed that the texture in NO is good in prediction of the observed BAU. In the case of IO, a very small parameter space agrees with the current cosmological limits on BAU. However, a small parameter space of B3 is in good agreement with the upper limits on the sum of the light neutrino masses. 
	
		\begin{figure}[H]
		\begin{center}
			\includegraphics[width=0.30\textwidth]{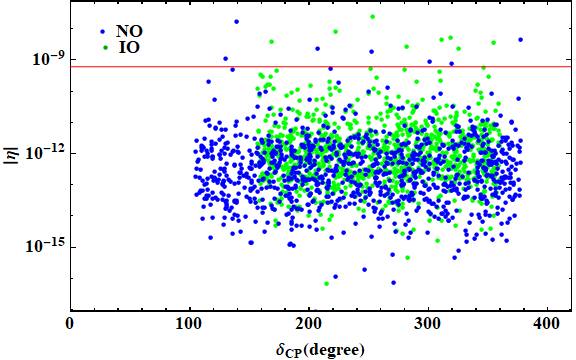}
			
		\end{center}
		\begin{center}
			\caption{Correlation between BAU and Dirac CP phase for the texture A1 in case of NO and IO.}
			\label{fig7}
		\end{center}
	\end{figure}
Fig \ref{fig7} represents the prediction of the model on $\delta_{CP}$ and baryon asymmetry of Universe for NO and IO in texture in one of the textures A1. It is evident that $\delta_{CP}$ lie within $(105-380)^{0}$ for NO and $(140-360)^{0}$ for IO.
	
	\begin{figure}[H]
		\begin{center}
			\includegraphics[width=0.30\textwidth]{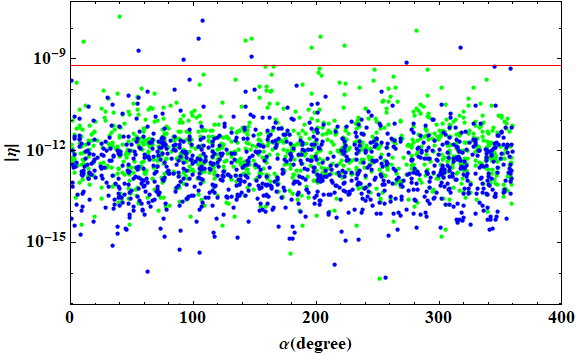}
			
		\end{center}
		\begin{center}
			\caption{Correlation between BAU and Majorana CP phase $\alpha$ for the texture A1 in case of both NO and IO.}
			\label{fig8}
		\end{center}
	\end{figure}
   Fig \ref{fig8} shows the correlation between baryon asymmetry and Majorana CP phase respectively for NO (Blue) and IO (Green) in texture A1. It has been observed that the texture predicts Majorana phase within $0-2\pi$. Similar results are obtained in the other five textures for $\delta_{CP}$ and Majorana CP phases.
   	\begin{figure}[H]
   	\begin{center}
   		\includegraphics[width=0.30\textwidth]{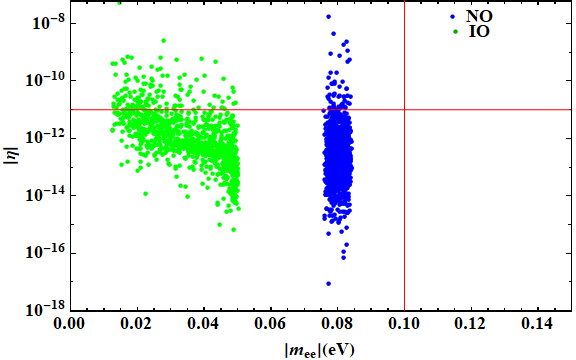}
   		
   	\end{center}
   	\begin{center}
   		\caption{Correlation between BAU and $m_{ee}$ for the texture A1 for both NO and IO.}
   		\label{fig9}
   	\end{center}
   \end{figure}
   The predictions of these textures on effective mass characterizing neutrinoless double beta decay have been shown in fig \ref{fig9}. It is observed that all these textures are in good agreement with the current experimental limits on the effective Majorana neutrino mass.
	\section{\label{sec:level8}Conclusion}
	In this work, we have studied the effect of two zero textures of Dirac mass matrix $M_{D}$ on leptogenesis in the framework of inverse seesaw ISS(2,3). The maximum allowed zeros in the structure of $M_{D}$ with the structures of $M_{N}$ and $\mu$ is two. We find the flavor symmetric origin of the six classes with $S_{4}$ discrete flavor symmetry. The $S_{4}$ flavor symmetry is further augmented with $Z_{4}\times Z_{3}$. The flavons get VEV after the spontaneous symmetry breaking and the different  VEVs can lead to different two zero structures of $M_{D}$. With the six allowed classes of two zero textures of $M_{D}$, we have studied the impact of texture zeros on the baryon asymmetry of the universe. It has been found that the textures A1, A3, and B3 in NO give rise to the observed baryon asymmetry. The other three textures are not in agreement with the current cosmological limit on BAU. Though texture A2 fails in explaining BAU in NO, yet it can lead to the observed BAU in IO. One can obtain the required amount of BAU with the textures A1 and B3 in IO also. Again, the texture A3, which is in good agreement in NO, can also yield correct amount of BAU in IO. However, all the textures are in good agreement with the current cosmological limit on the sum of the light neutrino masses. In our previous work \cite{Gautam:2021vhf}, it is observed that the textures A2 and B1 are not suitable from the cosmology data on dark matter phenomenology. In this work also it has been observed that the textures A2 and B1 are not able to yield the observed baryon asymmetry. So, in ISS(2,3) framework the two textures A2 and B1 can be discarded from the latest cosmology data. However, these textures are in good agreement with the neutrino data. As can be seen from \cite{Gautam:2021vhf}, the texture B2 in NO can account for good dark matter phenomenology, but it fails to explain BAU as can be observed from our results. Also B2 in IO is weak in explaining both dark matter and BAU. It can also be concluded that the texture A1 in ISS(2,3) is in good agreement with the latest cosmology data on dark matter as well as BAU for both NO and IO. Similarly, the textures A3 and B3 in NO can give good DM phenomenology along with the correct value of observed BAU.
	\begin{acknowledgments}
	NG acknowledges Department of Science and Technology (DST), India(grant DST/INSPIRE Fellowship/2016/IF160994) for the financial assistantship. 
	\end{acknowledgments}
\bibliographystyle{paper}
\bibliography{tex}

\end{document}